\documentclass[12pt]{iopart}
\usepackage{graphicx}% Include figure files
\begin{document}

\title[$K^*$(892) Production in Au+Au and pp Collisions
at $\sqrt{s_{NN}}$=200GeV at STAR]{$K^*$(892) Production in Au+Au
and pp Collisions at $\sqrt{s_{NN}}$=200GeV at STAR}

\author{Haibin Zhang\dag\ for the STAR Collaboration
\footnote[3]{See reference \cite{1} for the full collaboration
list.} }

\address{\dag\ Yale University, Physics Department, P.O.Box 208121, New Haven,
CT 06520-8121, USA, Email: zhang@hepmail.physics.yale.edu}

\begin{abstract}

Mid-rapidity $K^{*0}$(892)$\rightarrow K\pi$ and
$K^{*\pm}$(892)$\rightarrow K_S^0\pi^{\pm}$ are measured in Au+Au
and pp collisions at $\sqrt{s_{NN}}$=200GeV using the STAR
detector at RHIC. The $K^{*0}$(892) mass is systematically shifted
at small transverse momentum for both Au+Au and pp collisions. The
$K^{*0}$(892) transverse mass spectra are measured in Au+Au
collisions at different centralities and in pp collisions.  The
$K^{*0}$(892) mean transverse momentum as a function of the
collision centrality is compared to those of identified $\pi^{-}$,
$K^{-}$ and $\bar{p}$. The $K^{*}/K$ and $\phi/K^{*}$ ratios are
compared to measurements in A+A, $pp$, $\overline{p}p$,
$e^{+}e^{-}$ collisions at various colliding energies. The physics
implications of these measurements are also discussed.

\end{abstract}


\submitto{\JPG}

%\maketitle

\section{Introduction}

Resonance particles, which have very short lifetime (few fm/c)
comparable to the time scale for the evolution of the hot-dense
matter formed in ultra-relativistic heavy-ion collisions, can be a
very unique tool to probe the dynamics and properties of the high
density matter~\cite{2,3,4}. In particular, the $K^{*}$(892) has a
lifetime of $\sim$4fm/c, and can be produced at the chemical
freeze-out stage. The $K^{*}$ short lifetime makes it possible for
the newly-formed $K^{*}$ to undergo a period of re-interaction in
the hadronic gas phase. A portion of $K^*$ may decay before the
kinetic freeze-out stage and their kaon and pion daughter
particles might be re-scattered by other particles in the hadron
gas. This effect of the $K^{*}$ daughter particles re-scattering
can destroy part of the overall $K^{*}$ signal. On the other hand,
kaon and pion particles in the hadron gas can regenerate $K^{*}$
through the so-called pseudo-elastic collisions~\cite{6}. This
regeneration effect can compensate for the rescattering effect.
Thus the measurement of $K^{*}$ yields and their centrality
dependence in heavy-ion collisions can provide information to
estimate the time between chemical and kinetic freeze-out in
relativistic heavy-ion collisions~\cite{5,6}.\\

In the strongly interacting matter at high temperature and high
densities, characteristics of the short-lived $K^{*}$(892)
resonance might be modified in the medium, with shifted mass,
broadened width and even significantly changed line shapes. In the
hadron gas, kaon and pion particles can regenerate $K^*$ signals
through $K\pi\rightarrow K^*\rightarrow K\pi$. This regeneration
channel can also interfere with the kaon and pion elastic
scattering channel through $K\pi\rightarrow K\pi$. Thus the $K^*$
meson may be modified not only due to this interference but also
due to the kaon and pion initial phase space
distributions~\cite{7}. In addition, dynamical interactions of the
$K^*$ meson with the surrounding matter may also cause the
modification of the $K^*$ mass, width and line shape~\cite{8,9}.
Even though the size of the system formed in pp collisions is
smaller than in Au+Au collisions, interactions that may modify the
$K^*$ resonance are also expected. Thus a measurement of
$K^{*}$(892) mass, width and line shape in Au+Au and pp collisions
can provide very interesting information on possible in-medium
effects.\\

\section{Data Analysis and Results}

Preliminary measurements on $K^{*0}$(892)$\rightarrow K\pi$ and
$K^{*\pm}$(892)$\rightarrow K_S^0\pi^{\pm}$ in Au+Au and pp
collisions at $\sqrt{s_{NN}}$=200GeV are presented. These
measurements were made at the Solenoidal Tracker at RHIC (STAR)
with the main detector Time Projection Chamber (TPC). In Au+Au
collisions, the minimum bias trigger was defined by coincidences
between two Zero Degree Calorimeters (ZDC). A scintillator Central
Trigger Barrel (CTB) was used to select central collision events.
In pp collisions, the minimum bias trigger was defined using
coincidences between two beam counters that measured the charged
particles multiplicity near beam rapidity. After requiring the
collision vertex to be within $\pm$50cm along the beam line, about
2M top 10$\%$ central triggered, 2M minimum bias triggered Au+Au
collision events and 6M minimum bias triggered pp collision events
were used in this analysis. The events from minimum bias Au+Au
collisions were divided in four centrality bins from the most
central to peripheral collisions: 0$\%$-10$\%$, 10$\%$-30$\%$,
30$\%$-50$\%$ and 50$\%$-80$\%$.\\

Through energy loss ($dE/dx$) in the TPC gas, charged pions and
kaons were identified. In the case of $K^{*0}$, charged kaons were
selected by requiring their $dE/dx$ to be within two standard
deviations (2$\sigma$) of the expected value. A looser $dE/dx$ cut
of 3$\sigma$ was used for charged pions. In the case of
$K^{*\pm}$, $K_S^0$ candidates were selected from their decay
vertex geometries via $K_S^0\rightarrow\pi^{+}\pi^{-}$. The
invariant mass was then calculated for each kaon and pion pair in
an event. The invariant mass distribution derived in this manner
was then compared to a reference distribution calculated using
uncorrelated kaons and pions from different events. In this
analysis, $K^{*0}$ and $\overline{K^{*0}}$ are added together due
to limited statistics. The term $K^{*0}$ refers to the average of
$K^{*0}$ and $\overline{K^{*0}}$ unless specified otherwise.\\

\begin{figure}
\centering
\includegraphics[height=10pc,width=15pc]{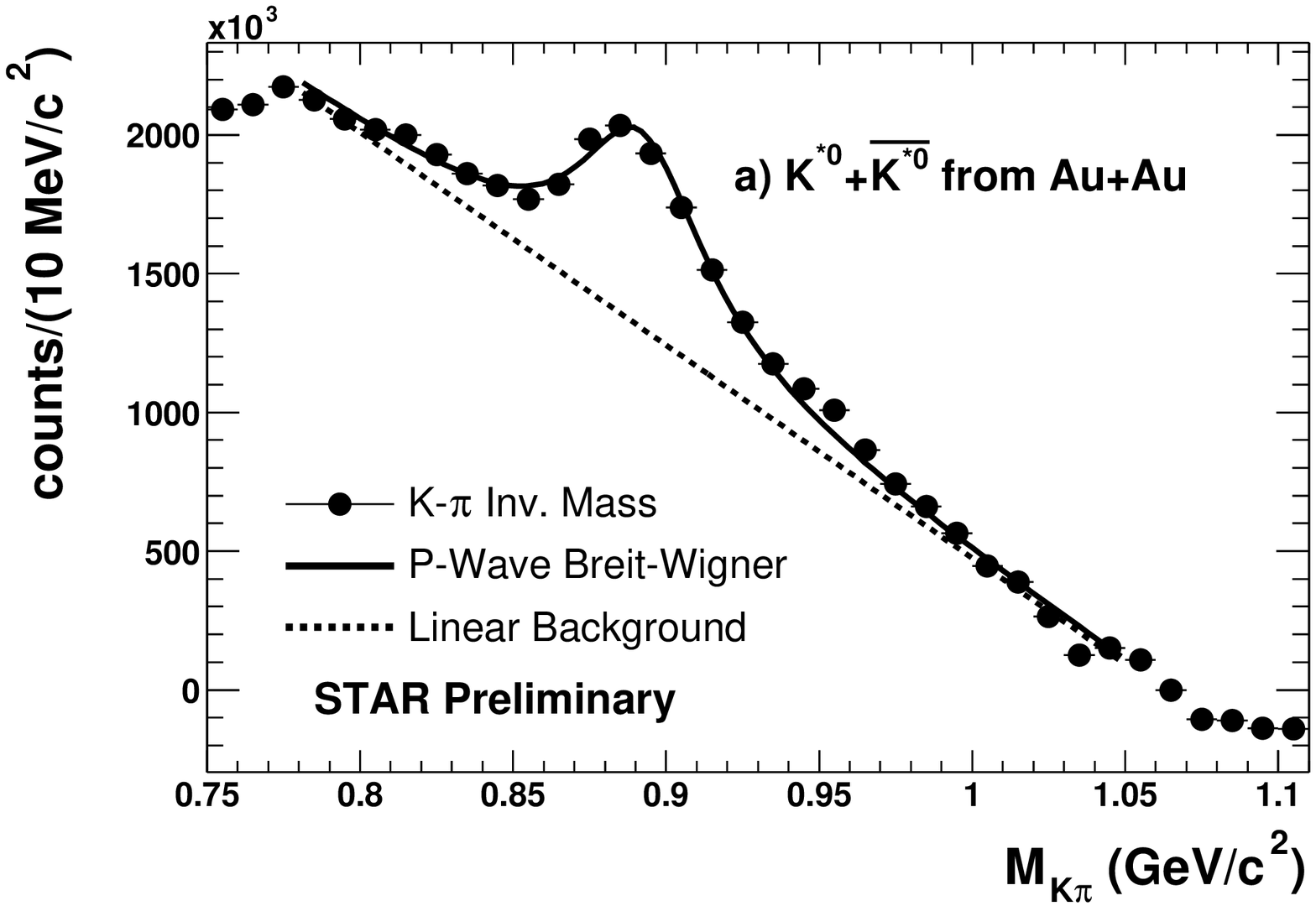}
\includegraphics[height=10pc,width=15pc]{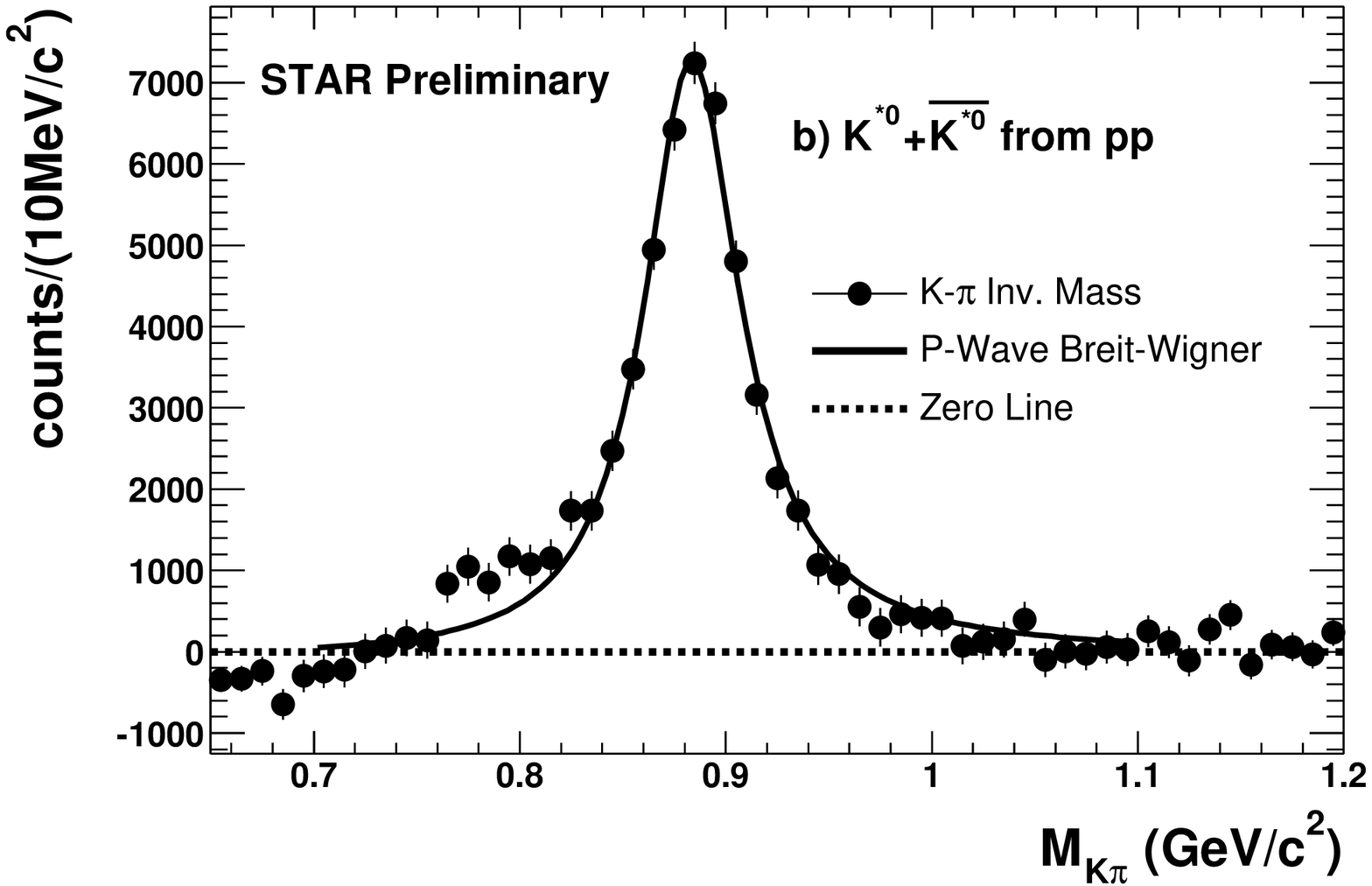}
\includegraphics[height=10pc,width=15pc]{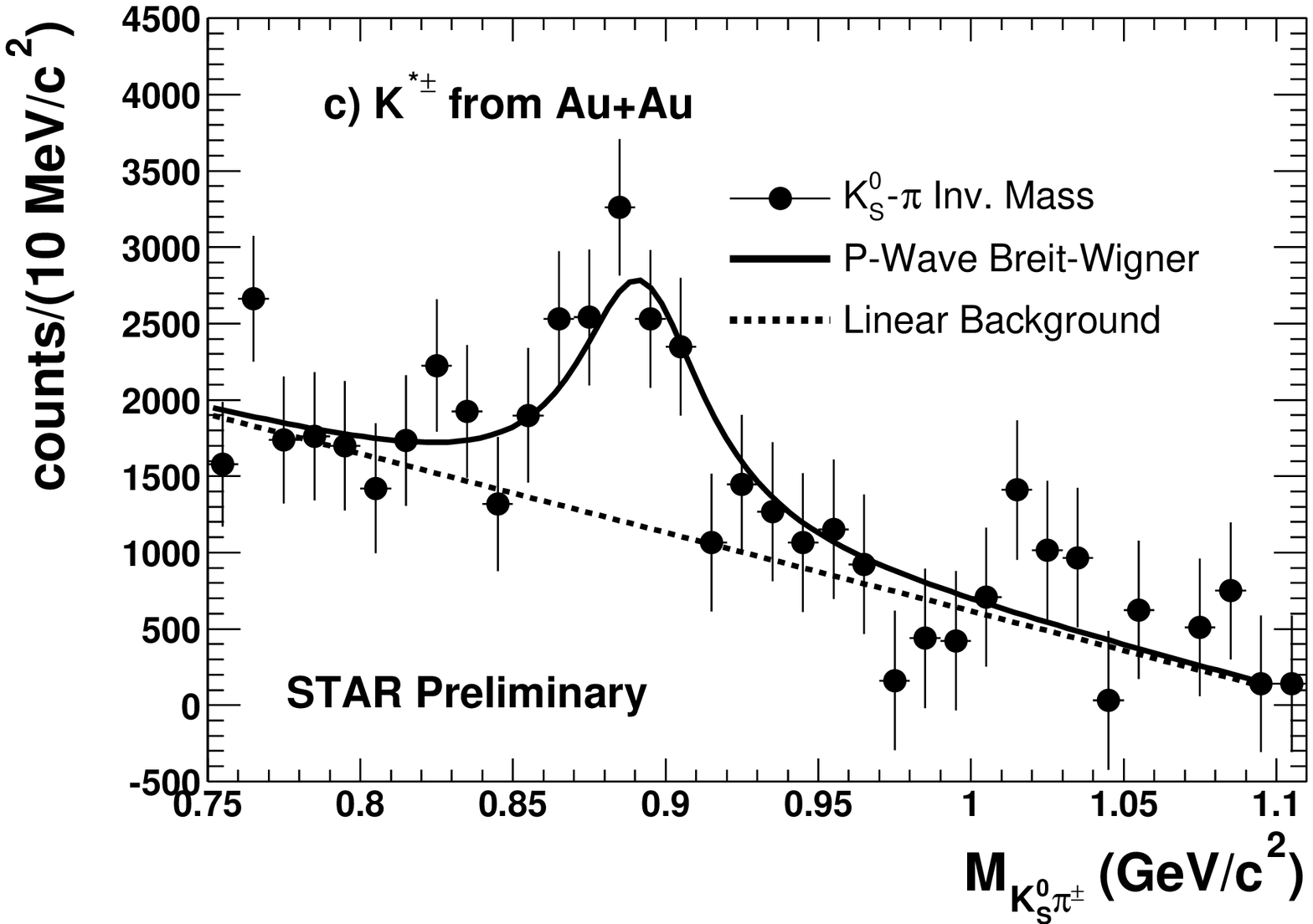}
\includegraphics[height=10pc,width=15pc]{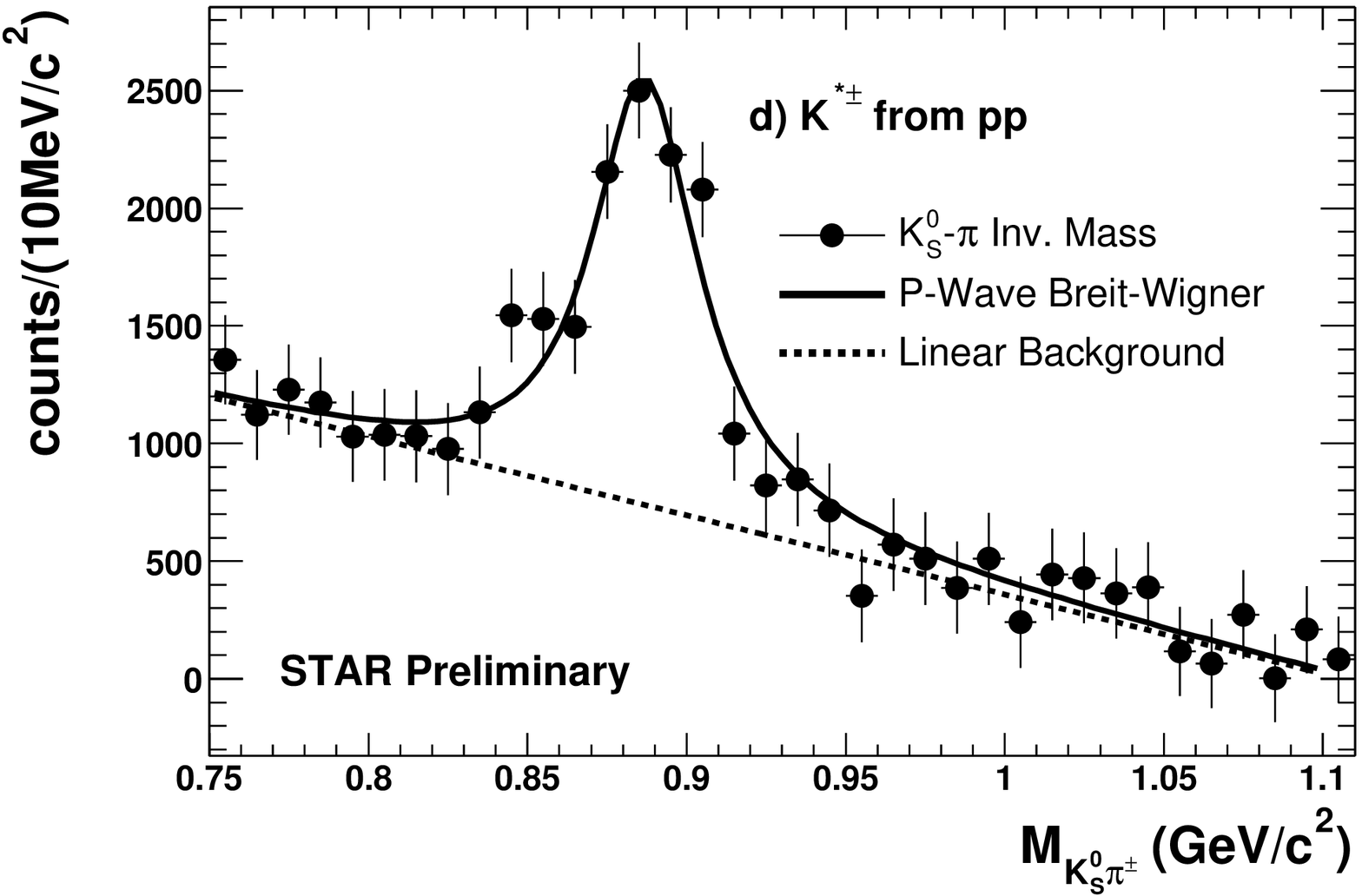}
\caption{The raw $K\pi$ invariant mass distribution after
subtraction of the mixed-event reference distribution for Au+Au
and pp collisions. a) $K^{*0}$ in top 10$\%$ central Au+Au
collisions; b) $K^{*0}$ in minimum bias pp collisions; c)
$K^{*\pm}$ in 50$\%$-80$\%$ hadronic cross section Au+Au
collisions; d) $K^{*\pm}$ in minimum bias pp collisions.}
\end{figure}

Figure 1 shows the $K\pi$ invariant mass distribution after
subtraction of the mixed-event reference distribution in Au+Au and
pp collisions. The invariant mass distributions are fit to a
combination of a p-wave Breit-Wigner function and a linear
function representing the background. In Figure 1 (a), there is a
large amount of residual background in the invariant mass
distribution after subtracting the reference distribution.
Reference~\cite{10} discussed possible sources of this residual
background. In addition, while mixing Au+Au collision events,
different events will have different reaction planes so that the
reference invariant mass distribution might have slightly
different shapes from the same event distribution. After
subtracting the reference distribution, this slight difference
will appear as part of the residual background~\cite{17}. In
Figure 1 (b), in order to precisely measure the $K^{*0}$ mass and
width as a function of transverse momentum in pp collisions, we
only selected the kaon candidates with momentum between 0.2 GeV/c
and 0.7 GeV/c to minimize the residual background. In Figure 1 (c)
and (d), $K_S^0$ candidates were first reconstructed via decay
vertex geometries requiring $\arrowvert M_{K_S^0}-M_{\pi^+\pi^-}
\arrowvert <$ 15MeV/c$^2$. Then, after pairing $K_S^0$ candidates
with charged pion candidates in same events and mixed events,
invariant mass distribution for the $K^{*\pm}$ were obtained after
reference distribution subtraction.\\

\begin{figure}
\centering
\includegraphics[height=11pc,width=15pc]{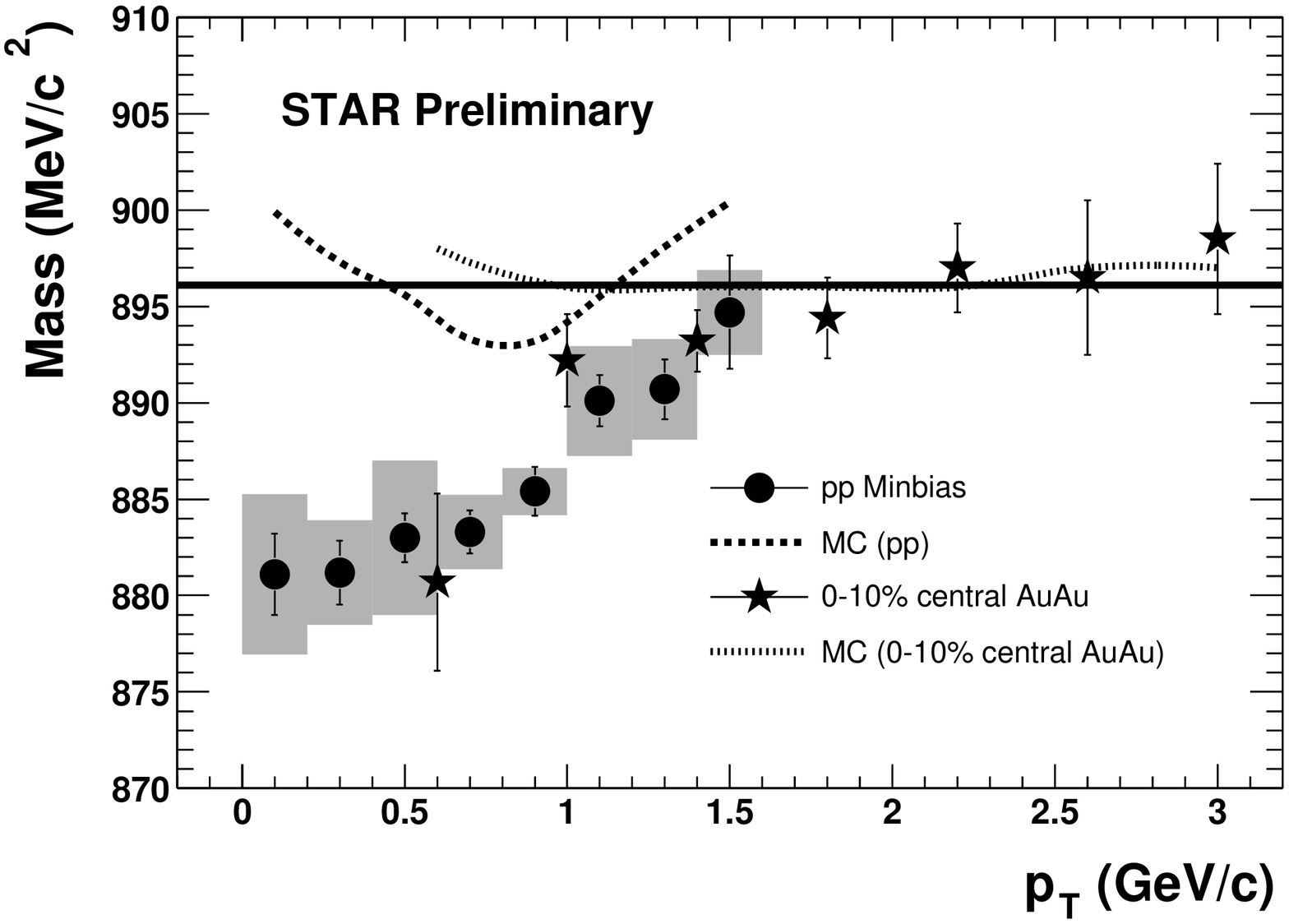}
\includegraphics[height=11pc,width=15pc]{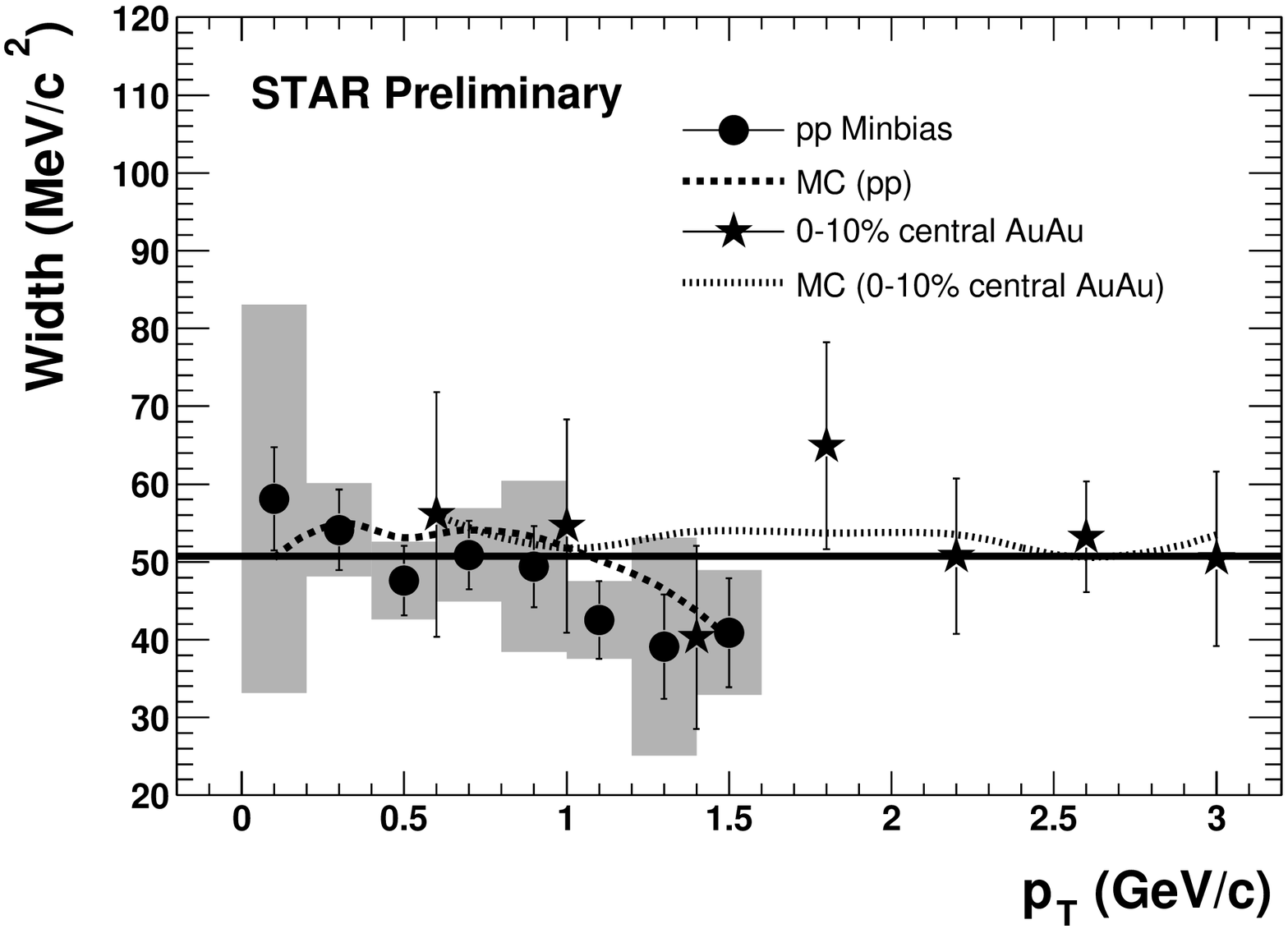}
\caption{$K^{*0}$ mass and width as a function of transverse
momentum for both Au+Au and pp collisions. The solid straight
lines stand for the standard $K^{*0}$ mass (896.1MeV/c$^2$) and
width (50.7MeV/c$^2$). The dashed curves represent the MC results
for $K^{*0}$ mass and width in pp collisions after considering
detector effects and kinematic cuts. The dotted curves represent
MC results in Au+Au collisions. The grey shadows are for
systematic uncertainties in pp.}
\end{figure}

The $K^{*0}$ mass and width as a function of transverse momentum
from Au+Au and pp collisions are shown in Figure 2. In pp
collisions, the $K^{*0}$ masses in low $p_T$ region are
systematically smaller than the Monte Carlo (MC) results which
account for detector effects and all kinematic cuts. The mass
shift is $p_T$ dependent, with the $K^{*0}$ mass increasing as a
function of $p_T$. In Au+Au collisions, the $K^{*0}$ mass shift is
also observed at low $p_{T}$. In the case of the $K^{*0}$ width,
there is no significant difference between the measured results
and the MC results in both Au+Au and pp collisions. As already
discussed in the introduction section, there are various physical
effects which may cause this $K^*$ mass shift. However, a soft
$K^*$ resonance (low $p_T$) is more likely to be modified in the
medium and thus a larger mass shift is expected for soft $K^*$
than for hard $K^*$ (high $p_T$). A similar mass shift for the
$\rho^0$(770) meson has also been observed in
both pp and Au+Au collisions\cite{11}.\\

\begin{figure}
\centering
\includegraphics[height=16pc,width=15pc]{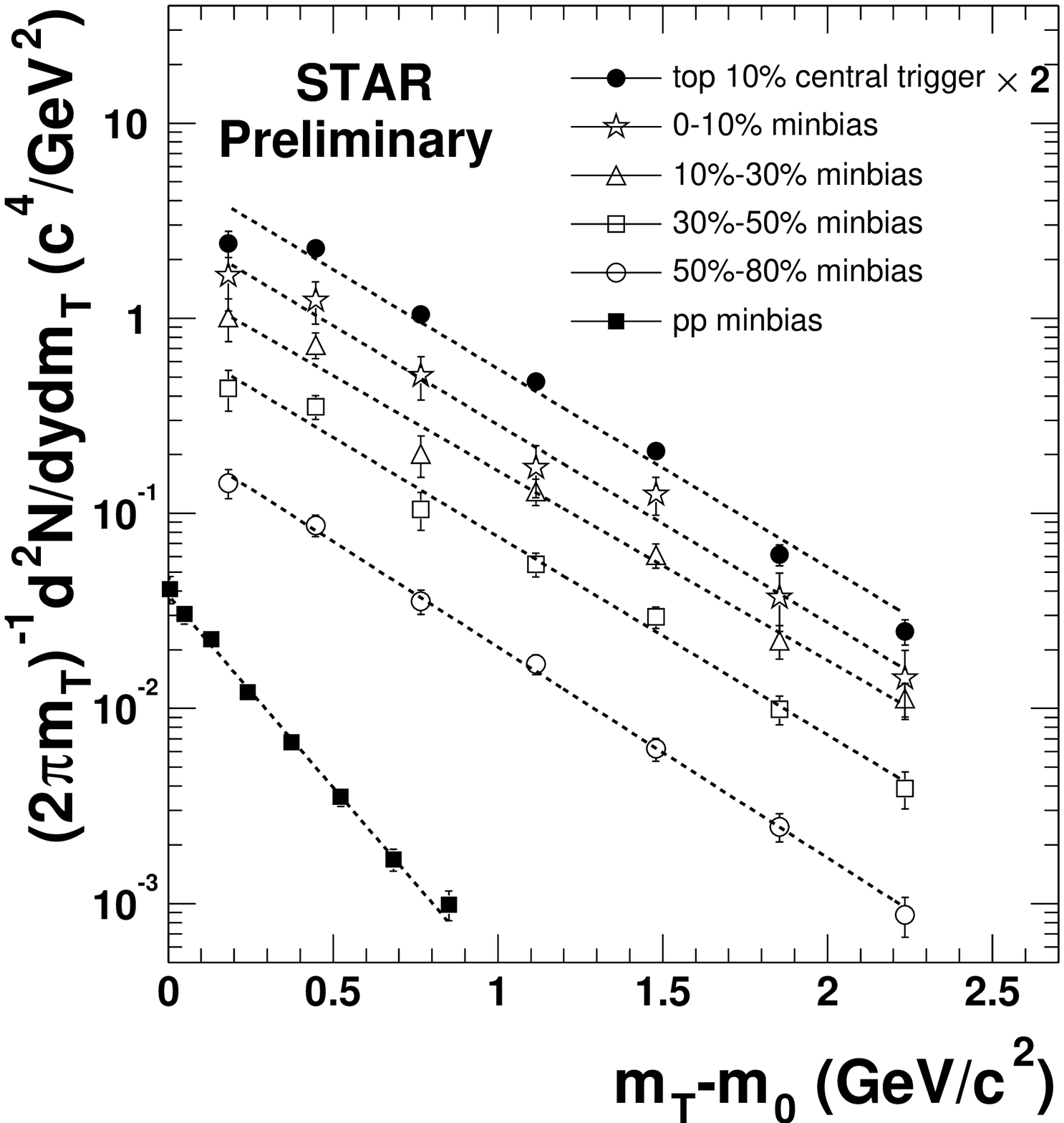}
\includegraphics[height=16pc,width=15pc]{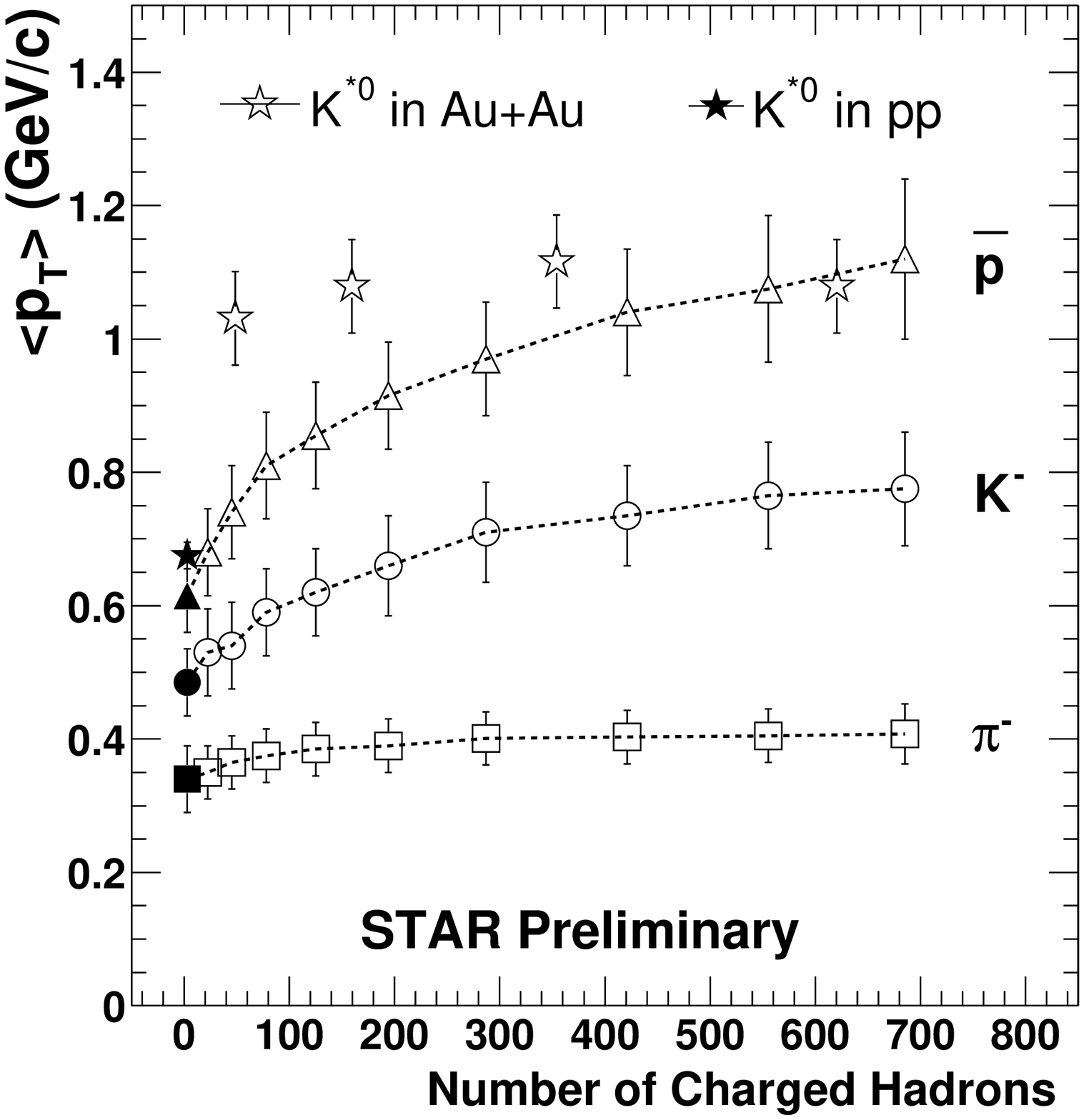}
\caption{$(K^{*0}+\overline{K^{*0}})/2$ $m_{T}$-distributions at
mid-rapidity ($|y|<0.5$) in Au+Au and pp collisions (left).
$K^{*0}$ mean $p_{T}$ as a function of number of charged hadrons
and compared to $\pi^-$, $K^-$ and $\overline{p}$ (right).}
\end{figure}

The detector acceptance and efficiency corrected $K^{*0}$
transverse mass ($m_T=\sqrt{p_T^2+m^2}$) spectra at mid-rapidity
($|y|<$0.5) are shown in the left plot of Figure 3. The spectra
are fit with exponential functions and $K^{*0}$ yields $dN/dy$ and
inverse slopes are extracted from the exponential fit. The
$K^{*0}$ $dN/dy$ increases from pp collisions to peripheral Au+Au
collisions and to central Au+Au collisions. The inverse slopes in
Au+Au collisions are systematically larger than that in pp
collisions. In the right plot of Figure 3, the $K^{*0}$ mean
$p_{T}$ as a function of number of charged hadrons is compared to
that of $\pi^-$, $K^-$ and $\overline{p}$~\cite{12}. In pp
collisions, the $K^{*0}$ mean $p_{T}$ is comparable to the mean
$p_{T}$ of $\overline{p}$. In Au+Au collisions, the $K^{*0}$ mean
$p_{T}$ are systematically larger than that in pp collisions. The
larger $K^*$ inverse slopes and mean $p_T$ in Au+Au collisions
than in pp collisions might be explained by the fact that in the
hadronic phase of Au+Au collisions, $K^*$ with higher $p_T$ are
more likely to escape and decay outside the fireball thus avoiding
the daughter particles' re-scattering. A low $p_T$ $K^*$ has more
chances to be destroyed by daughter particles' re-scattering in
the hadron gas medium in Au+Au collisions.\\

\begin{figure}
\centering
\includegraphics[height=16pc,width=15pc]{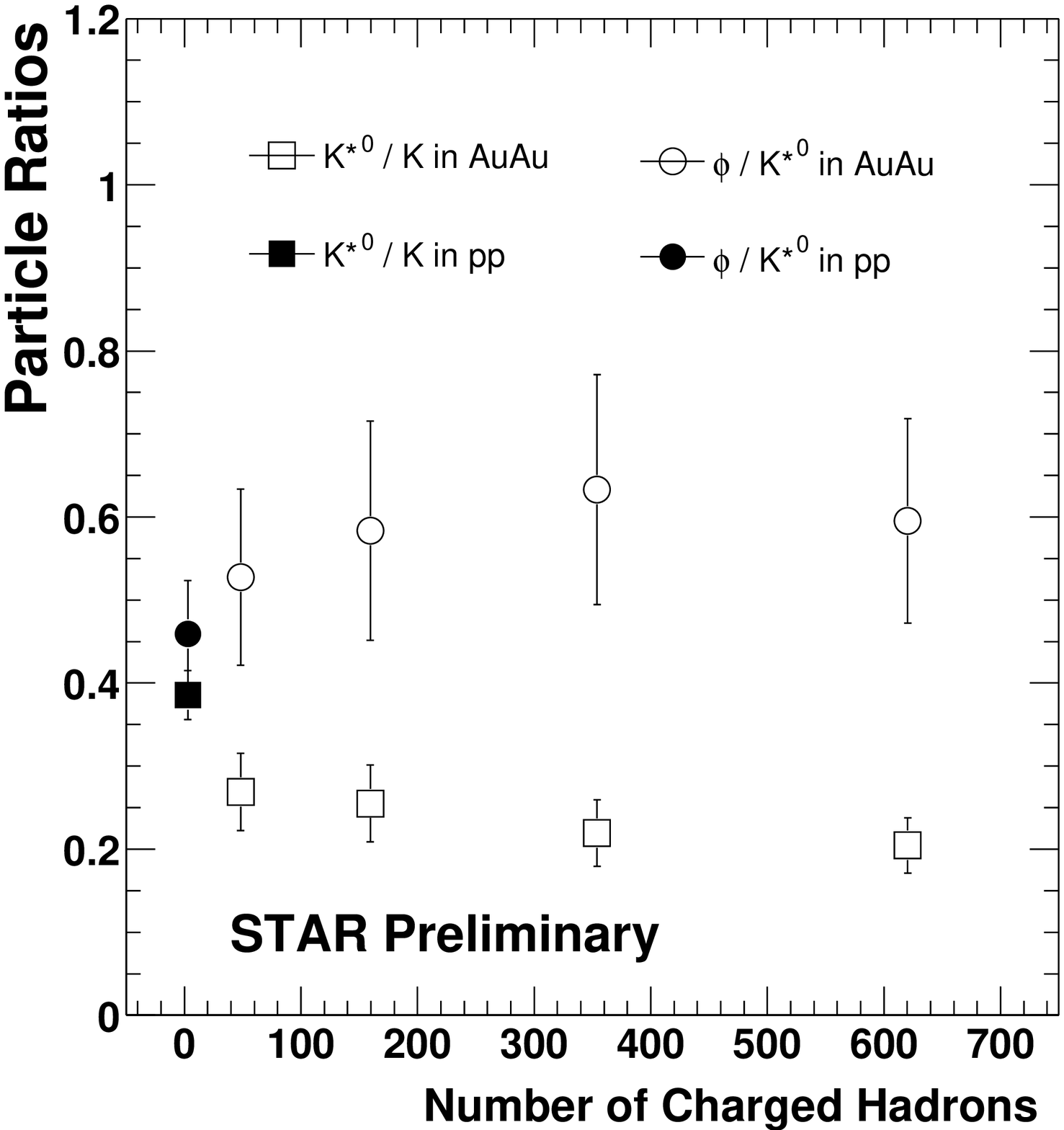}
\includegraphics[height=17pc,width=17pc]{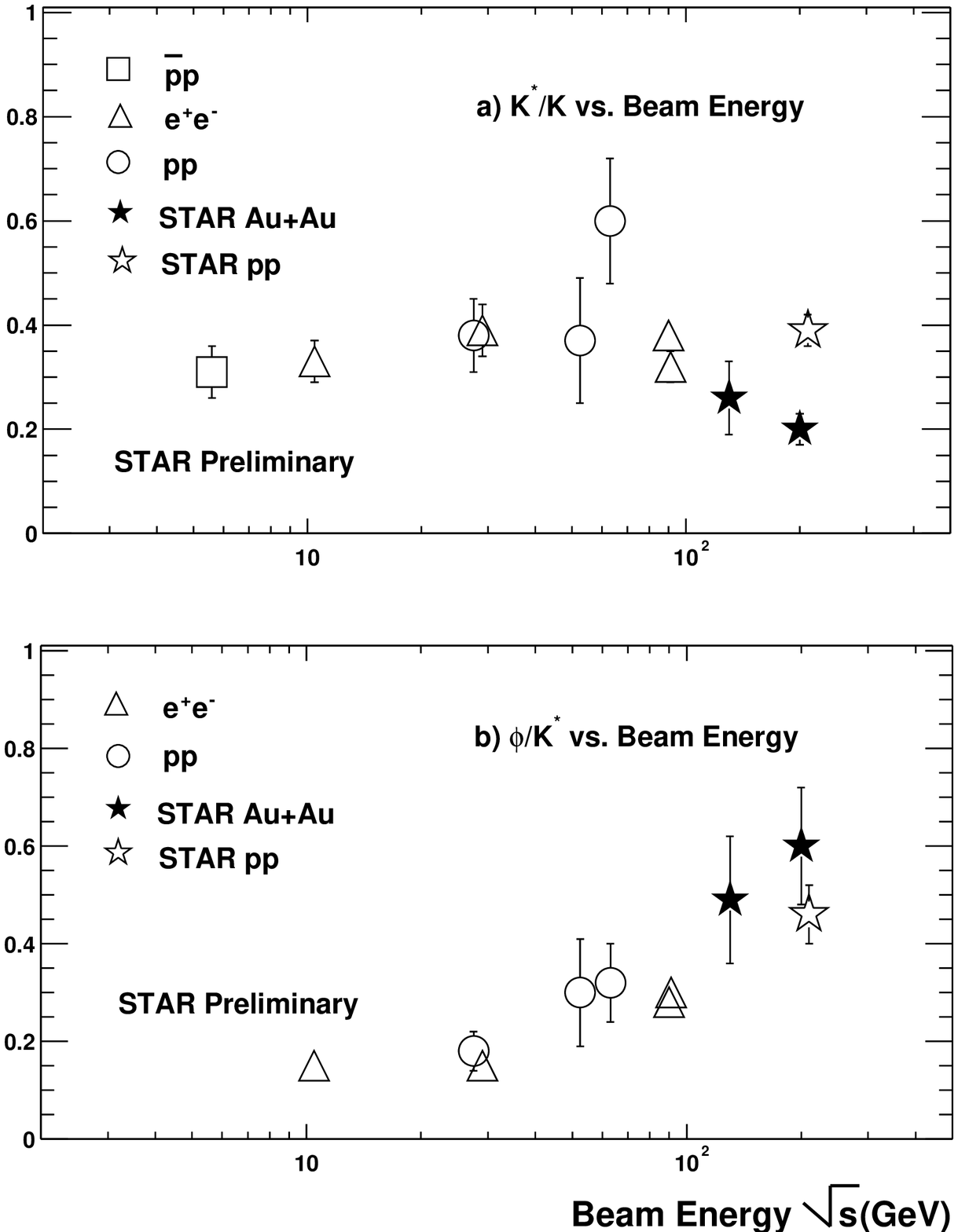}
\caption{$K^{*}/K$ and $\phi/K^{*}$ ratios as a function of number
of charged hadrons for both pp and Au+Au collisions (left).
$K^{*}/K$ and $\phi/K^{*}$ ratios compared to different collision
systems ($e^+e^-$~\cite{14}, $\overline{p}p$~\cite{15} and
$pp$~\cite{16}) at various collision energies (right).}
\end{figure}

$K^{*}$ and $K$ mesons of the same charge have identical quark
content and only differ in mass and spin so that the $K^{*}/K$
ratio might reflect the fireball evolution conditions from
chemical to kinetic freeze-out. Figure 4 shows that the $K^*/K$
ratios in Au+Au collisions are significantly reduced compared to
the ratio in pp collisions. This ratio reduction may indicate that
in Au+Au collisions, $K^*$ daughter particles' re-scattering might
destroy more $K^*$ than the amount of $K^*$ which are regenerated
by the $K\pi$ interactions in the hadron gas. The upper right
panel of Figure 4 shows the $K^*/K$ ratio as a function of the
colliding energy from different collision systems. The $K^*/K$
ratio from elementary collisions does not change noticeablly in
the presented $\sqrt{s}$ region. However, the ratio in Au+Au
collisions is significantly smaller. The $K^*$ and $\phi$ mesons
have a small mass difference and $\Delta S$=1. Thus, the
$\phi/K^{*}$ ratio might also be a good signature to study the
strangeness enhancement effect in ultra-relativistic heavy ion
collisions. The bottom right panel of Figure 4 shows that the
$\phi/K^*$ ratio increases as a function of the colliding energy.
We may need more statistics to identify any possible $\phi/K^*$
ratio differences between Au+Au collisions and pp collisions at
$\sqrt{s_{NN}}$=200GeV. In this analysis, $K$ results are
from~\cite{12} and $\phi$ results are from~\cite{13}, $K^*$
results for Au+Au collisions at $\sqrt{s_{NN}}$=130GeV are
from~\cite{10}.\\

\section{Summary}

In summary, we have measured the $K^{*0}$(892) and $K^{*\pm}$(892)
production from Au+Au and pp collisions at $\sqrt{s_{NN}}$=200GeV
at STAR. Possible in-medium dynamical effects might have modified
the $K^{*}$ mass line shape and thus systematically smaller
$K^{*}$ masses are observed at the low $p_{T}$ region in pp and
Au+Au collisions. In Au+Au collisions, the $K^{*}$ daughter
particles' re-scattering effect is dominant over the regeneration
effect. This picture is consistent with our measurements that the
$K^*$ inverse slopes and mean $p_{T}$ in Au+Au are larger than in
pp and
the $K^*/K$ ratios in Au+Au are smaller than in pp.\\


\begin{thebibliography}{9}
\bibitem{1}H. Caines, these proceedings.
\bibitem{2}R. Rapp \textit{et al.}, Adv. Nucl. Phys. 25, 1 (2000).
\bibitem{3}R. Rapp \textit{et al.}, Phys. Rev. C 63, 054907
(2001).
\bibitem{4}J. Schaffner-Bielich, Phys. Rev. Lett. 84, 3261 (2000).
\bibitem{5}G. Torrieri and J. Rafelski, hep-ph/0112195.
\bibitem{6}M. Bleicher \textit{et al.}, QM02 proceedings.
\bibitem{7}R. Longacre, paper in preparation.
\bibitem{8}E. Shuryak and G. Brown, Nucl.Phys. A717 (2003) 322-335.
\bibitem{9}E. Shuryak, hep-ph/0304145.
\bibitem{10}C. Adler \textit{et al.}, Phys. Rev. C 66, 061901(R) (2002).
\bibitem{11}P. Fachini, these proceedings.
\bibitem{12}O. Barannikova and F. Wang, QM02 proceedings.
\bibitem{13}J. Ma, these proceedings.
\bibitem{14}H. Albrecht \textit{et al.}, Z. Phys. C 61 (1994)1;
M. Derrick \textit{et al.}, Phys. Lett. B 158 (1985) 519; K. Abe
\textit{et al.}, Phys. Rev. D 59 (1999) 052001; Y. J. Pei
\textit{et al.}, Z. Phys. C 72 (1996) 39.
\bibitem{15}J. Canter \textit{et al.}, Phys. Rev. D 20 (1979)
1029.
\bibitem{16}T. Akesson \textit{et al.}, Nucl. Phys. B203, 27
(1982); D. Drijard \textit{et al.}, Z. Phys. C 9, 193 (1981); M.
Aguilar-Benitez \textit{et al.}, Z. Phys. C 50, 405 (1991).
\bibitem{17}L. Gaudichet, STAR note SN446.
\end{thebibliography}
\end{document}